\documentstyle[nato,amsmath,amssymb,psfig]{crckapb} 

\begin{opening}
\title{Time at the origin of the Universe: \protect\\
       fluctuations between two possibilities}

\author{V. Dzhunushaliev$^1$}
\institute{Affiliation \\
           Kyrgyz-Russian Slavic University, 
           Kievskaya Str, 44, 720000, Bishkek, Kyrgyzstan \\
           e-mail : dzhun@hotmail.kg}

\runningtitle{Time at the origin of the Universe: ... }

\end{opening}

\begin{document}


\begin{abstract}
A variation of Hawking's idea about Euclidean origin of a nonsingular 
birth of the Universe is considered. It is assumed that near to zero 
moment $t = 0$ fluctuations of a metric signature are possible. 
\end{abstract}

\section{Introduction}

The time in the modern Lorentz - invariant physics is connected 
with a metric signature. The spacetime metric for any space 
can be written as 
\begin{equation}
  ds^2 = g_{\mu\nu} dx^\mu dx^\nu = 
  e^{\bar{a}} e^{\bar{b}} \eta_{\bar{a}\bar{b}} = 
  \left( h^{\bar{a}}_\mu dx^\mu \right) 
  \left( h^{\bar{b}}_\nu dx^\nu \right) \eta_{\bar{a}\bar{b}}
\label{1}
\end{equation}
where $g_{\mu\nu} = h^{\bar{a}}_\mu h^b_\nu \eta_{\bar{a}\bar{b}}$ 
is the metric, 
$e^{\bar{a}} = h^{\bar{a}}_\mu dx^\mu$ is 1-form, 
$h^{\bar{a}}_\mu$ is vier-bein, 
$a$ is vier-bein index, $\mu$ is the coordinate index and 
$\eta_{\bar{a}\bar{b}}$ is the metric signature 
$\eta_{\bar{a}\bar{b}} = diag\{ \sigma,1,1,1 \}$. 
An undefined number $\sigma$ can be $+1$ for the Euclidean 
space and $-1$ for the Lorentzian spacetime. 
We see that the difference between Euclidean and Lorentzian spacetimes 
is connected with the sign of $\sigma = \eta_{\bar{0}\bar{0}} = \pm 1$. For 
$\eta_{\bar{0}\bar{0}} = +1$ we say that there is the Euclidean space and for 
$\eta_{\bar{0}\bar{0}} = -1$ the Lorentzian spacetime. 
\par
\begin{figure}
\framebox{
\begin{minipage}[b]{.40\linewidth}
\psfig{figure=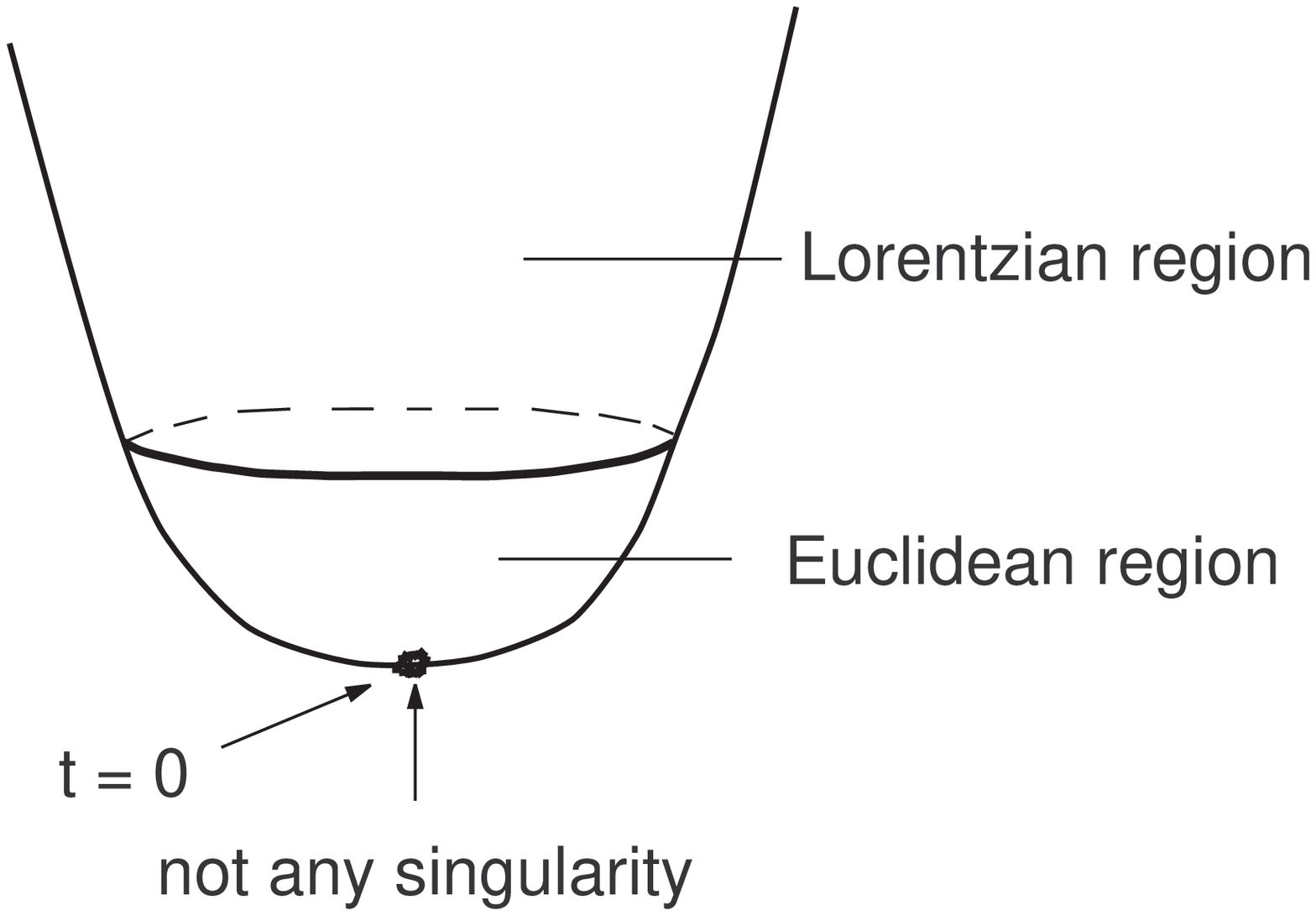,width=5cm,height=3cm}
\caption{Hawking's nonsingular Universe with the Euclidean region at the 
origin. }
\label{fig1}
\end{minipage}
}\hfill
\framebox{
\begin{minipage}[b]{.40\linewidth}
\psfig{figure=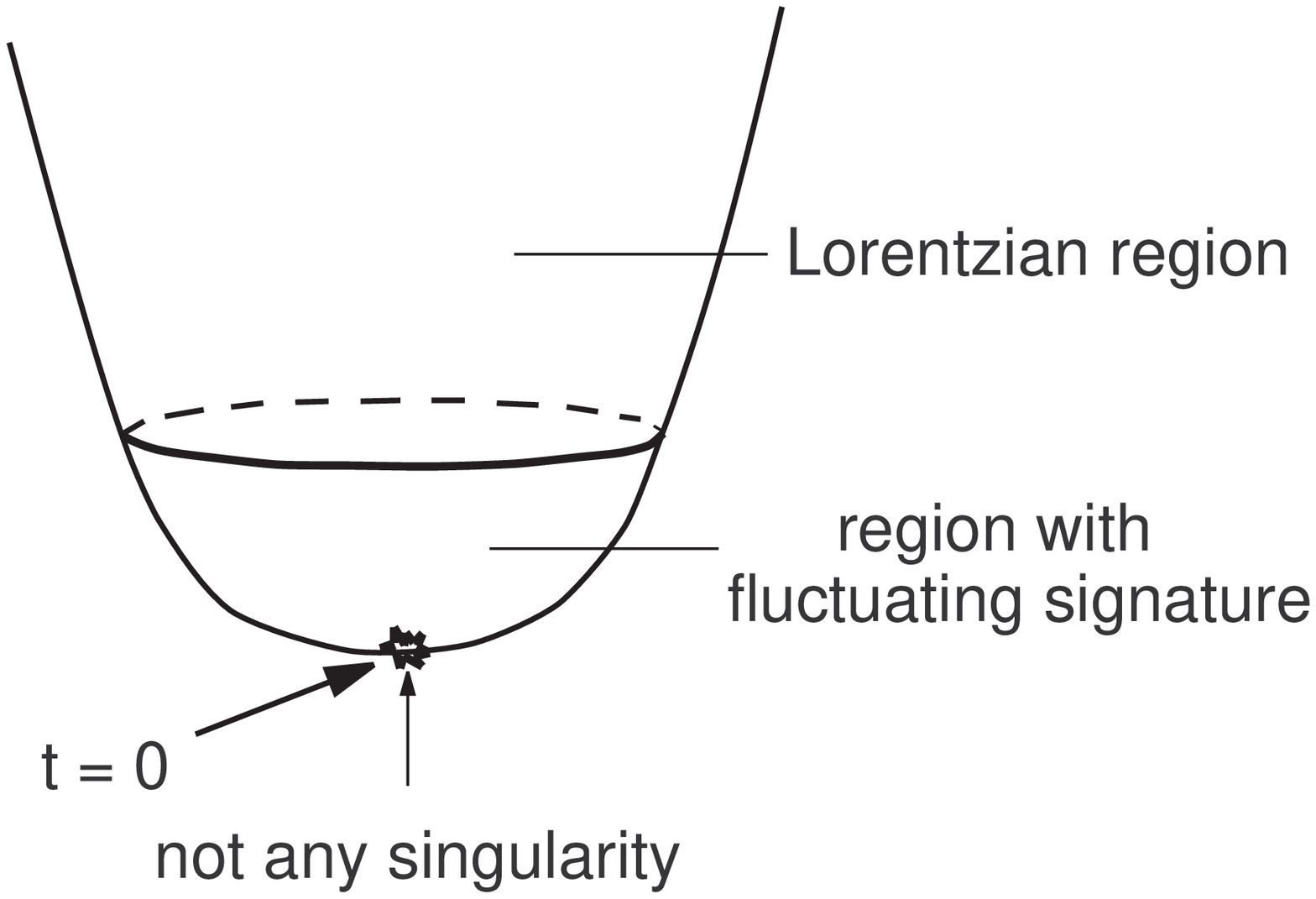,width=5cm,height=3cm}
\caption{At the origin of Universe there is a region with fluctuations 
of the metric signature.}
\label{fig2}
\end{minipage}
}
\end{figure}

 
It easy to see  that the metric $g_{\mu \nu}$ 
has two different degrees of freedom : vier-bein 
$h^{\bar{a}}_\mu$ and the metric signature $\eta_{\bar{a}\bar{b}}$. 
$h^{\bar{a}}_\mu$ can be determined 
from Einstein's equations but for the metric signature $\eta_{\bar{a}\bar{b}}$ we have 
not any dynamical equations. 
We put in the metric signature $\eta_{\bar{a}\bar{b}}$ by hand into 
Einstein's equations (in the vier-bein formalism). Of coarse, 
we can determine the true value of $\eta_{\bar{a}\bar{b}}$ from experiments : 
in our Universe $\eta_{\bar{a}\bar{b}} = (-1,1,1,1)$. But on the quantum 
level (on the Planck level) we can assume that $\eta_{\bar{a}\bar{b}}$ 
is fluctuating quantity 
\begin{equation}
  diag \left( -1, 1, 1, 1 \right)
  \leftrightarrows
  diag \left( +1, 1, 1, 1 \right) .
\label{4}
\end{equation}
Cosmological solutions of Einstein's equations with ordinary matter 
have almost without exception a cosmological singularity. The existence 
of such singularity is one of the most significant problem in the 
modern physics. There are various approaches to the solution of this 
problem. Hawking's point of view \cite{hawking} is that at the origin of time 
an Euclidean space emerged from Nothing and the metric signature near 
to zero moment $t = 0$ has the Euclidean value
$\eta_{ab} = diag \left( +1,1,1,1 \right)$ 
but after short time interval ($\approx t_{Pl}$) the signature undergoes 
a quantum jump to the Lorentzian value
$\eta_{ab} \rightarrow diag \left( -1,1,1,1 \right)$ 
Another words : \textit{in order to kill singularity we must kill 
the time.} Thus we have the following picture for Hawking's nonsingular 
Universe (see, Fig.\ref{fig1}). 
The idea presented here is that near to zero moment there are quantum 
fluctuations between Euclidean and Lorentzian signatures and after 
short time ($\approx t_{Pl}$) the fluctuations cease and the metric comes 
to the state with the definite (Lorentzian) metric signature 
(see, Fig.2). 
\par 
We can assume that in quantum gravity can be various 
sort of quantum fluctuations 
\begin{enumerate}
    \item 
    The fluctuations of the metric.
    \item 
    The topology fluctuations. This phenomenon is known as a hypothesized 
    spacetime foam. 
    \item 
    The fluctuations of the metric signature.
    \item $\ldots$
\end{enumerate} 
Here I will consider only the third kind of quantum gravitational 
fluctuations. One can say that two different approaches to the problem 
of the metric signature fluctuations are possible :
\begin{enumerate}
    \item 
    One can solve the Einstein's equations with undefined $\sigma = \pm 1$ 
    and then we will have fluctuating quantity $\sigma$ in the solution. 
    \item 
    Another approach is that every Einstein's equation (for example, 
    $R_{00} - \frac{1}{2}g_{00} R = 0$) with $\sigma = + 1$ and 
    $\sigma = - 1$ can have different probability.
\end{enumerate}    
We can interpret the second approach as follows : in fact Einstein's 
equations are some algorithm for calculations of the metric in whole space. 
In this approach we have an algorithm some parts of which can fluctuate 
(see, Fig.3). 
\begin{figure}
\begin{center}
\framebox{
\psfig{figure=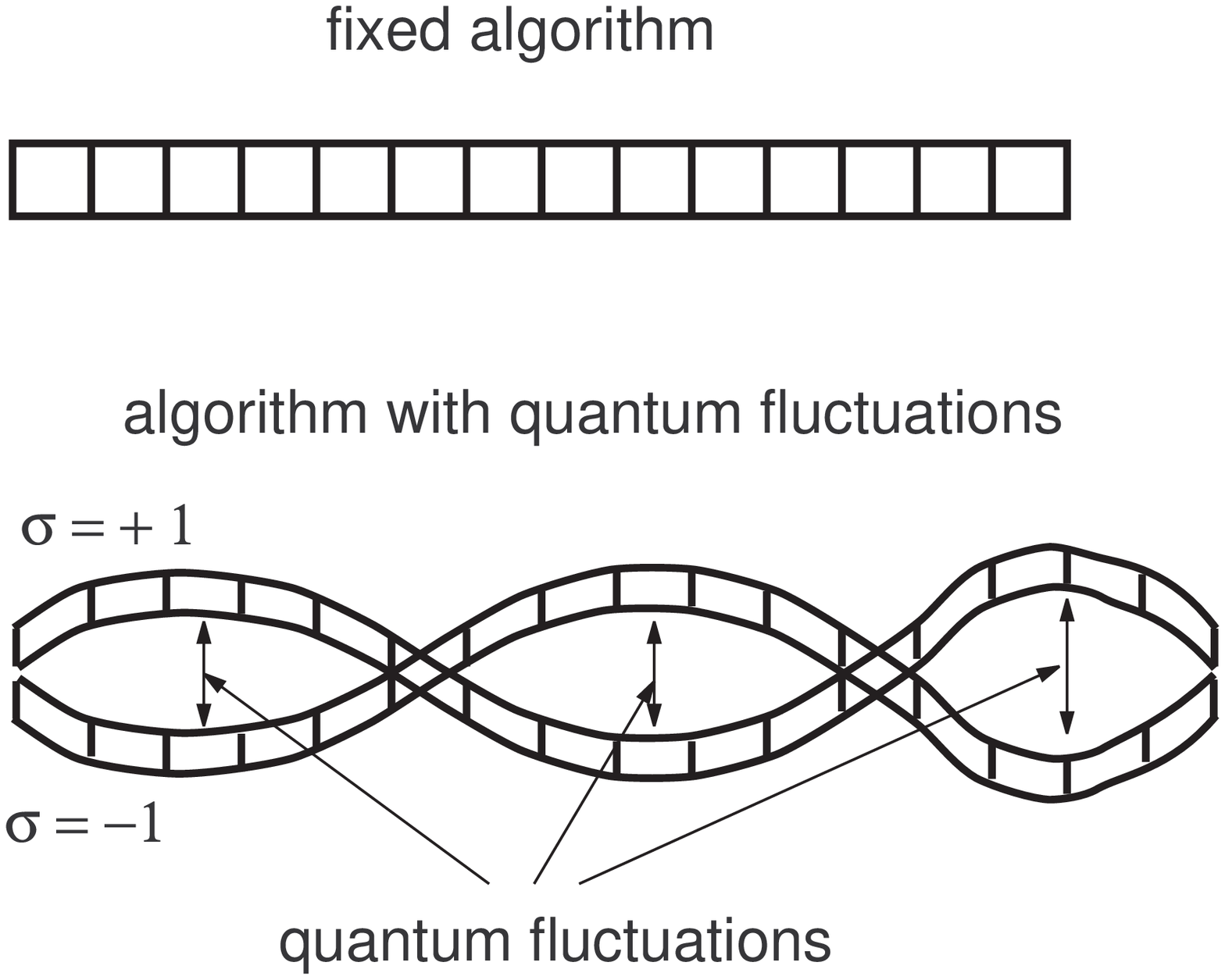,width=5cm,height=3cm}}
\caption{A fluctuating algorithm. }
\end{center}
\label{fig4}
\end{figure}
\par 
We assume that 
the probability of every version of every Einstein's equation is 
connected with a ``complexity'' of the equation. 
\textbf{What is it the ``complexity'' ?} Of coarse intuitively it is clear. 
Here we can recall Einstein's statement that
``Everything should be made as simple as possible, but not simpler.''
Our proposal is that the ``complexity'' is connected
with Kolmogorov's ideas on algorithmic complexity (AC).
In this approach any physical system ({\it e.g.} the Universe)
can be thought of in terms of an algorithm. The longer and
more complex the algorithm, the less likely it is for such
a system to appear. In particular Universes
with different physical laws (field equations) are described
by different algorithms. The length of these algorithms then
affects the probability that this Universe with a certain set of
physical laws will fluctuate into existence.
In our case we will search such combination of the Einstein's equations 
with the different $\sigma = \pm 1$ so that the solution will be the simplest. 
At first we will give an exact definition for the complexity. 

\section{Kolmogorov's algorithmic complexity}

The mathematical definition for algorithmic complexity (AC) is 
\par
\textit{
The algorithmic complexity $K(x\mid y)$ of the  object $x$  for a 
given object $y$ is the minimal length of the ``program'' $P$
that is written as a sequence of  the  zeros  and  ones
which allows us to construct $x$ starting from $y$:
\begin{equation}
K(x\mid y) = \min_{A( P,y)=x} l(P)
\label{ac1}
\end{equation}
$l(P)$ is length of the  program $P$; $A(P,y)$  is  the
algorithm  for calculating an object $x$, using  the  program $P$,
when the object $y$ is given.} 

\section{The 5D Fluctuating Universe}

Here I would like to consider the scenario where at
the origin of the Universe fluctuations
between Euclidean and Lorentzian metrics occur \cite{dzh} \cite{dzhsin}. 
We start with a vacuum 5D Universe with the metric 
\begin{equation}
\begin{split}
ds^2_{(5)} = \sigma dt^2 + b(t)\left (d\xi + 
\cos \theta d\varphi \right )^2 + a(r)d\Omega ^2_2 + \\  
r_0^2 e^{2\psi (t)}\left [d\chi - \omega (t) \left (d\xi + 
\cos \theta d\varphi \right ) \right ]^2 
\end{split}
\label{fluct60}
\end{equation}
here $\sigma = \pm 1$ for the Euclidean and 
Lorentzian signatures respectively. 
According to the appropriate theorem the multidimensional 
metric in Eq. \eqref{fluct60} has the following electromagnetic potential
\begin{equation}
A = \omega (t) \left (d\xi + 
\cos \theta d\varphi \right ) = 
\frac{\omega}{\sqrt b} e^{\bar 1}
\label{fluct100}
\end{equation}
which yields an electrical field $E_{\bar 1}$ and a magnetic field 
$H_{\bar 1}$ like
\begin{eqnarray}
E_{\bar 1} = F_{\bar 0 \bar 1} & = & \frac{\dot \omega}{\sqrt b}
\label{fluct120}\\
H_{\bar 1} = \frac{1}{2}\epsilon_{1\bar j\bar k}F^{\bar j \bar k} & = & 
- \frac{\omega}{a}
\label{fluct130}
\end{eqnarray}
The 5D, vacuum Einstein equations resulting from
Eq. \eqref{fluct60} are
\begin{eqnarray}
G_{\bar 0\bar 0} \propto 
2\frac{\dot b \dot \psi}{b} + 4 \frac{\dot a \dot \psi}{a} + 
2\frac{\dot a \dot b}{ab} + \frac{\dot a^2}{a^2} + 
\sigma \left (\frac{b}{a^2} - \frac{4}{a}\right ) + && 
\nonumber \\ 
r_0^2e^{2\psi}\left (\sigma H_{\bar 1}^2 - E_{\bar 1}^2 
\right ) & = & 0 , 
\label{fluct140a}\\
G_{\bar 1\bar 1} \propto 4\ddot\psi + 4{\dot\psi}^2 + 
4\frac{\ddot a}{a} + 4\frac{\dot a\dot\psi}{a} + 
\sigma\left (3\frac{b}{a^2} - \frac{4}{a} \right ) - && 
\nonumber \\
\frac{{\dot a}^2}{a^2} + 
r_0^2e^{2\psi}\left (\sigma H_{\bar 1}^2 - E_{\bar 1}^2 
\right ) & = & 0 , 
\label{fluct140b}\\
G_{\bar 2\bar 2} = G_{\bar 3\bar 3} \propto 
4\ddot\psi + 4{\dot\psi}^2 + 
2\frac{\ddot b}{b} + 2\frac{\dot b\dot\psi}{b} - 
\frac{\dot b^2}{b^2} + 2\frac{\ddot a}{a} + && 
\nonumber \\
2\frac{\dot a\dot\psi}{a} + 
\frac{\dot a\dot b}{ab} - \frac{\dot a^2}{a^2} - 
\sigma\frac{b}{a^2} - 
r_0^2e^{2\psi}\left (\sigma H_{\bar 1}^2 - E_{\bar 1}^2 
\right ) & = & 0 , 
\label{fluct140c}\\
R_{\bar 5\bar 5} \propto \ddot \psi + {\dot\psi}^2 + 
\frac{\dot a\dot\psi}{a} + \frac{\dot b\dot\psi}{2b} + 
\frac{r_0^2}{2}e^{2\psi}\left (
\sigma H_{\bar 1}^2 + E_{\bar 1}^2
\right ) = 0 , &&
\label{fluct140d}\\
R_{\bar 2\bar 5} \propto \ddot \omega + \dot\omega
\left (\frac{\dot a}{a} - \frac{\dot b}{2b} + 3\dot\psi
 \right )
-\sigma \frac{b}{a^2}\omega & = & 0 
\label{fluct140e}
\label{fluct140}
\end{eqnarray}
The whole of algorithm is the concatenation of 
$G_{00} \times G_{11} \times \ldots \times R_{25}$ equations but in 
our approach every piece of the algorithm  (for example $G_{00}$) 
can fluctuate (here 
$G_{\bar A\bar B} = R_{\bar A\bar B} - \frac{1}{2}\eta_{\bar A\bar B}R$ 
is the Einstein tensor). Our basic assumption is that 
\textit{at the Planck scale there can exist regions where
quantum fluctuations between Euclidean and Lorentzian metric signatures
occur}. There are two copies of the classical equations \eqref{fluct140}:
one with $\sigma = +1$ and another with $\sigma = -1$. The basic question 
under our assumption
is how to calculate the relative probability for each pair of equations
from \eqref{fluct140} (the ones with $\sigma = +1$ versus the ones with
$\sigma = -1$).
\par
We will define the probability for each pair of equations in terms
of the AC of each pair. We can diagrammatically
represent the fluctuations between the Euclidean and Lorentzian
versions of Einstein's equations in the following way
\begin{equation}
\begin{array}{ccc}
\sigma = +1 & \longleftrightarrow & \sigma = -1
\\
& \Downarrow  & 
\\
\left(G^+\right)_{\bar 0\bar 0} 
& 
\longleftrightarrow
& 
\left(G^-\right)_{\bar 0\bar 0} 
\\
\left(G^+\right)_{\bar 1\bar 1} 
& 
\longleftrightarrow
& 
\left(G^-\right)_{\bar 1\bar 1}
\\
\left(G^+\right)_{\bar 2\bar 2} 
& 
\longleftrightarrow
& 
\left(G^-\right)_{\bar 2\bar 2}
\\
\left(G^+\right)_{\bar 3\bar 3} 
& 
\longleftrightarrow
& 
\left(G^-\right)_{\bar 3\bar 3}
\\
\left(R^+\right)_{\bar 5\bar 5} 
& 
\longleftrightarrow
& 
\left(R^-\right)_{\bar 5\bar 5}
\end{array}
\label{fluct150}
\end{equation}
The signs $\pm$ indicates if the equation belongs to the
Euclidean or Lorentzian mode.
Expression \eqref{fluct150} sums up the idea that treating
$\sigma$ as a quantum quantity leads to quantum fluctuations between
the classical equations:
$\left(R^+\right)_{\bar A\bar B} \leftrightarrow
\left(R^-\right)_{\bar A\bar B}$ or $\left(G^+\right)_{\bar A\bar B}
\leftrightarrow \left(G^-\right)_{\bar A\bar B}$.
The probability connected with each pair of equations
($R^\pm_{\bar A\bar B}$  or $G^\pm_{\bar A\bar B}$) 
\textit{is determined by the AC of each equation.}

\paragraph{Fluctuation $\left(R^+\right)_{\bar 2\bar 5} 
\longleftrightarrow \left(R^-\right)_{\bar 2\bar 5}$.} 

The $R_{\bar 2\bar 5}$ equation in the Euclidean and Lorentzian modes is 
respectively 
\begin{eqnarray}
\ddot \omega + \dot \omega\left ( \frac{\dot a}{a} - 
\frac{\dot b}{2b} + 3\dot \psi \right )
-\frac{b}{a^2} \omega = 0 ,
\label{fluct160a} \\
\ddot \omega + \dot \omega\left ( \frac{\dot a}{a} - 
\frac{\dot b}{2b} + 3\dot \psi \right )
+ \frac{b}{a^2} \omega = 0 .
\label{fluct160b}
\label{fluct160}
\end{eqnarray}
Let us consider the $\psi = 0$ case (below we 
will see that this is consistent with the
$R_{\bar 5\bar 5}$ equation). It is easy to see that Eq.
\eqref{fluct160a} can be deduced from the instanton condition 
\begin{equation}
E^2_{\bar 1} = H^2_{\bar 1} \qquad or \qquad 
\frac{\omega}{a} = \pm \frac{\dot\omega}{\sqrt b}
\label{fluct180}
\end{equation}
The second equation \eqref{fluct160b} does not have
a similar simplification via the instanton condition
\eqref{fluct180}. Based of this simplification
from a second order equation \eqref{fluct160a} to a first order equation
\eqref{fluct180} we consider the Euclidean equation \eqref{fluct160a}
simpler from an algorithmic point of view than the Lorentzian 
equation \eqref{fluct160b}. To a first, rough approximation 
we can take the probability of the Euclidean 
mode as $p^+_{25} \approx 1$ and for the Lorentzian mode as $p^-_{25}
\approx 0$. 

\paragraph{Fluctuation $\left(R^+\right)_{\bar 5\bar 5} 
\longleftrightarrow \left(R^-\right)_{\bar 5\bar 5}$.} 

The $R_{\bar 5\bar 5}$ equation in the Euclidean and Lorentzian modes is 
respectively 
\begin{eqnarray}
\ddot \psi + {\dot\psi}^2 + 
\frac{\dot a}{a}\dot\psi + \frac{\dot b}{b}\dot\psi + 
\frac{r_0^2}{2}e^{2\psi}\left (
H^2_{\bar 1} + E^2_{\bar 1}
\right ) =  0 ,
\label{fluct200a}\\
\ddot \psi + {\dot\psi}^2 + 
\frac{\dot a}{a}\dot\psi + \frac{\dot b}{b}\dot\psi + 
\frac{r_0^2}{2}e^{2\psi}\left (
-H^2_{\bar 1} + E^2_{\bar 1}
\right ) =  0 .
\label{fluct200b}
\end{eqnarray}
The Lorentzian mode \eqref{fluct200b} has a trivial solution
\begin{equation}
\psi = 0
\label{fluct220}
\end{equation}
provided the instanton condition ({\it i.e.} $H^2_{\bar 1}
= E^2_{\bar 1}$) holds and describes the ``frozen'' 5 coordinate. 
Thus for this equation we take
the Lorentzian mode as having a smaller AC, and
in the contrast with the previous subsection,
the Lorentzian mode has the greater probability. Again
to a first, rough approximation the
probability of the Euclidean mode is $p^+_{55} \approx 0$ and 
consequently for the Lorentzian mode $p^-_{55} \approx 1$ . 

\paragraph{Fluctuation $\left(G^+\right)_{\bar 1\bar 1} 
\longleftrightarrow \left(G^-\right)_{\bar 1\bar 1}$ 
and $G^+_{\bar 2\bar 2} \longleftrightarrow G^-_{\bar 2\bar 2}$. }

Taking into account \eqref{fluct220} we can write these
equations as 
\begin{eqnarray}
4\frac{\ddot a}{a} + 
\sigma\left (3\frac{b}{a^2} - \frac{4}{a} \right ) - 
\frac{{\dot a}^2}{a^2} + 
r_0^2 \left (\sigma H_{\bar 1}^2 - E_{\bar 1}^2 
\right ) & = & 0 ,
\label{fluct240a}\\
2\frac{\ddot b}{b}  - 
\frac{\dot b^2}{b^2} + 2\frac{\ddot a}{a} + 
\frac{\dot a\dot b}{ab} - \frac{\dot a^2}{a^2} - 
\sigma\frac{b}{a^2} - r_0^2
\left (\sigma H_{\bar 1}^2 - E_{\bar 1}^2
\right ) & = & 0 .
\label{fluct240b}
\end{eqnarray}
For the Euclidean mode ($\sigma = +1$) with the instanton condition 
\eqref{fluct180} one can have $b = a$ (an isotropic Universe) which
reduces the two equations of \eqref{fluct240a} - \eqref{fluct240b} 
to \textit{only} one equation
\begin{equation}
4\frac{\ddot a}{a} - \frac{{\dot a}^2}{a^2} - \frac{1}{a} = 0 .
\label{fluct250}
\end{equation}
For the Lorentzian mode ($\sigma = -1$) $b \ne a$ 
(an anisotropic Universe) there are still two equations 
\begin{eqnarray}
4\frac{\ddot a}{a} - 
\left (3\frac{b}{a^2} - \frac{4}{a} \right ) - 
\frac{{\dot a}^2}{a^2} - 
r_0^2 \left (H_{\bar 1}^2 + E_{\bar 1}^2 
\right ) & = & 0 ,
\label{fluct260a}\\
2\frac{\ddot b}{b}  - 
\frac{\dot b^2}{b^2} + 2\frac{\ddot a}{a} + 
\frac{\dot a\dot b}{ab} - \frac{\dot a^2}{a^2} + 
\frac{b}{a^2} + 
r_0^2 \left (H_{\bar 1}^2 + E_{\bar 1}^2 
\right ) & = & 0 ,
\label{fluct260b}
\end{eqnarray}
Thus under the instanton condition \eqref{fluct180} and $\psi =0$ we
find that the Euclidean mode \eqref{fluct250} effectively
reduces to one equation which corresponds to an
isotropic Universe; the Lorentzian mode \eqref{fluct260a} - \eqref{fluct260b} 
still has two equations which describe an
anisotropic Universe. Thus we assign the Euclidean mode the smaller
AC and as for the previous equations make the rough approximation
$p^+_{11} \approx 1$ for the Euclidean mode,
$p^-_{11} \approx 0$ for the Lorentzian mode.

\paragraph{Fluctuation $\left(G^+\right)_{\bar 0\bar 0} 
\longleftrightarrow \left(G^-\right)_{\bar 0\bar 0}$.} 

The equation  $G^\pm_{\bar 0\bar 0} = 0$ has the
following form
\begin{equation}
2\frac{\dot b \dot \psi}{b} + 4 \frac{\dot a \dot \psi}{a} + 
2\frac{\dot a \dot b}{ab} + \frac{\dot a^2}{a^2} + 
\sigma \left (- \frac{4}{a} + \frac{b}{a^2} \right ) + 
r_0^2e^{2\psi}\left (\sigma H_{\bar 1}^2 - E_{\bar 1}^2 
\right ) = 0
\label{fluct270}
\end{equation}
Assuming all the previous conditions (the instanton condition,
$\psi = 0$, and $b = a$) the Euclidean mode equations become
\begin{equation}
\frac{\dot a^2}{a^2} - \frac{1}{a} = 0 
\label{fluct280} 
\end{equation}
while the Lorentzian mode equations become
\begin{equation}
3\frac{\dot a^2}{a^2} + 3 \frac{1}{a} -
r_0^2 \left (H_{\bar 1}^2 + E_{\bar 1}^2 
\right )= 0 .
\label{fluct290}
\end{equation}
The instanton condition again implies that the Euclidean
mode has a smaller AC. Thus to a first, rough 
approximation we take $p^+_{00} \approx 1$ 
and $p^-_{00} \approx 0$. 

\section{Mixed system of the equations.} 

Under the approximation where the probability associated
with each of the equations in \eqref{fluct140} is
$p \approx 0$ or $1$ the \textit{mixed} system of equations
which describe a Universe \textit{fluctuating between Euclidean
and Lorentzian modes}
\begin{eqnarray}
\frac{\dot a^2}{a^2} - \frac{1}{a} & = & 0 , 
\label{fluct300a}\\
\dot \omega & = & \pm \frac{\omega}{\sqrt a} ,
\label{fluct300b}\\
4\frac{\ddot a}{a} - \frac{{\dot a}^2}{a^2} 
- \frac{1}{a} & = & 0 .
\label{fluct300c}
\end{eqnarray}
here $b = a$, $\psi = 0$ and the instanton condition are all
assumed to hold. This system of mixed Euclidean and Lorentzian
equations has the following simple solution
\begin{eqnarray}
a & = & \frac{t^2}{4} ,
\label{fluct310a}\\
\omega & = & t^2 .
\label{fluct310b}
\end{eqnarray}
We can interpret a small piece (with linear size of the Planck length 
$\approx l_{Pl}$) of our model 5D Universe as a quantum birth 
of the regular 4D Universe. 

\section{Conclusions}

In this talk we have considered the possibility that Nature
can have changing the physical laws. We have postulated
that the dynamics of this changing may be connected with the AC
of a particular set of laws. This leads to the proposition that
\textit{an object with a  smaller AC has a greater probability to
fluctuate into existence.}
Some physical consequences that can results from this hypothesized
fluctuation of physical laws at the Planck scale are:
\begin{itemize}
    \item 
    the birth of the Universe with a fluctuating metric signature;
    \item 
    the transition from a fluctuating metric signature to Lorentzian one;
    \item 
    ``frozen'' 5$^{th}$ dimension as a consequence of this transition.
\end{itemize}

\end{document}